\documentclass[aip,amsmath,amssymb,reprint,]{revtex4-1}

\usepackage{graphicx}
\usepackage{dcolumn}
\usepackage{bm}

\usepackage[utf8]{inputenc}
\usepackage[T1]{fontenc}
\usepackage{mathptmx}
\usepackage{etoolbox}
\usepackage{xcolor}
\usepackage{ulem}

\makeatletter
\def\@email#1#2{%
 \endgroup
 \patchcmd{\titleblock@produce}
  {\frontmatter@RRAPformat}
  {\frontmatter@RRAPformat{\produce@RRAP{*#1\href{mailto:#2}{#2}}}\frontmatter@RRAPformat}
  {}{}
}%
\makeatother
\begin{document}

\preprint{AIP/123-QED}

\title[Three robust temperature-drift compensation strategies for a MEMS gravimeter]{Three robust temperature-drift compensation strategies for a MEMS gravimeter}

\author{Victor M. Valenzuela}
\affiliation{ 
Facultad de Ciencias Físico Matemáticas, Universidad Autónoma de Sinaloa, Sinaloa, 80013, MX}

\author{Daniel Teran}
\author{Alejandro Sandoval}
\affiliation{Instituto de Física, Universidad Autónoma de San Luis Potosí, San Luis Potosí, 78295, MX
}
\author{Eduardo Gomez}
\altaffiliation[]{Author to whom correspondence should be addressed: egomez@ifisica.uaslp.mx}{}
\affiliation{Instituto de Física, Universidad Autónoma de San Luis Potosí, San Luis Potosí, 78295, MX
}

\author{John A. Franco-Villafañe}
\affiliation{CONAHCYT - Instituto de Física, Universidad Autónoma de San Luis Potosí, San Luis Potosí, 78295, MX
}

\author{Jesus J. Alcantar-Peña}
\author{Juan Ponce-Hernandez}
\affiliation{Centro de Ingeniería y Desarrollo Industrial, Querétaro, 76125, MX}

\date{\today}

\begin{abstract}
Gravimeters fabricated with MEMS suffer from temperature-dependent drifts in their long-term stability. We analyze the thermal contributions to the signal, and we propose three mechanisms to mitigate their effects. The first one uses materials that fulfill the condition $\alpha_E=-2 \alpha$, where the thermal expansion is canceled by the temperature variation of the Young's modulus. The second one uses the thermal expansion to introduce a compression that compensates the variation in the force of the spring. In the third one, the expansion compensates the displacement of the proof mass in the sensor, rather than the force. The three mechanisms are robust since they only depend on the temperature of the sensor itself. 
\end{abstract}

\maketitle

\section{Introduction}
Recent progress in precision gravimetry has found extensive applications both in fundamental physics and technology. They appear in proposals to test Non-Newtonian gravity at small scales \cite{Tino21,Moore21,Westphal21}, quantum gravity \cite{Amelino-Camelia14,Arndt14,Carney19-2} and dark matter detection \cite{Carney19,Carney21}. Applications include inertial navigation \cite{Yazdi98,Shaeffer13}, monitoring of volcanic activity \cite{Greco12,Carbone17}, seismometry \cite{Imanishi04,Zou14,Montagner16}, exploration of underground resources like water or oil \cite{Arndt14,Jacob10,VanCamp17,Chaffaut22}, and even monitoring climate changes \cite{vanDam17,Wang18GRACE}. 

The gravimeters are divided into absolute and relative. The first ones include the FG5-X gravimeter \cite{FG5} and those using cold \cite{Kasevich91,Peters99,Bongs19,Hu13} or ultra-cold atoms \cite{Aguilera14,Abend16}. Unfortunately they are expensive and not very practical for field applications due to their weight and complexity, even with the most recent versions \cite{Freier16,Bidel18,Menoret18,Wu19,Bidel20,Heine20,Muquans22}.

Gravimeters based on micro-electromechanical-system (MEMS) have been gaining importance due to their reliability, low cost and reduced power consumption \cite{Middlemiss16,Middlemiss17,Tang19,Wang20,Lu21}. They measure the displacement $z$ or the oscillation frequency $\omega_m$ of a proof mass $m$ attached to a spring with elastic constant $k$. Their drifts can be mitigated by having absolute gravimeters nearby for re-calibration \cite{Francis98,Sugihara08,Crossley13,Hector15,VanCamp16,Abrykosov19,Aspelmeyer14}. The damping $\gamma$ determines the ultimate sensitivity and long-term stability, and having a high mechanical quality factor $Q_m=\omega_m / \gamma$ provides good isolation from environmental noise sources \cite{Aspelmeyer14,Romero20,Sementilli22}. In some gravimeters the mechanical spring is replaced by a magnetically or optically levitated mass \cite{Prothero68,Goodkind99,Timberlake19,Monteiro20,Lewandowski21,Qvarfort18,iGrav22}.

The gravity acceleration is obtained from
\begin{equation}
    g=\frac{k}{m} z=\omega_m^2 z.
    \label{eq:g_relative}
\end{equation}
MEMS gravimeters reach a noise floor of about $1$  $\mu Gal/\sqrt{Hz}$ \cite{Wu20,Xu21}. In the AC (resonant) configuration, $g$ is measured through a frequency shift $\Delta \omega_m$. This method has a high resolution and large range (no displacement limitation) \cite{Su05,Tocchio11,Mustafazade20}, but it requires high quality factors, is more prone to failure by fatigue, and the temperature-dependent frequency drifts limit the long-term stability \cite{Zhang15,Wang18MEMS,Yin19,Fang21,Langfelder11}. The DC (static) configuration is easier to implement since $g$ is obtained from the mass displacement, monitoring the vertical position of a slit located in the test mass. It has good enough resolution and range \cite{Tang19,Wu20,Wu21} to perform state-of-the-art precision gravity measurements and there is much less risk of damage by fatigue since passive structures do not experience sufficient mechanical stress cycles to trigger fatigue \cite{Tsuyoshi12}.

The long-term stability still requires improvement in MEMS gravimeters, particularly on the temperature sensitivity \cite{Belwanshi22}. Strategies in this direction include temperature stabilization \cite{Jha08,Salvia10,Lee12,Liu18}, passive compensation through doping  \cite{Hajjam12,Samarao12,Langfelder11}, materials with opposite thermal coefficients \cite{Melamud09,Tabrizian13}, differential structure design \cite{Comi10,Park14,Mustafazade20}, mode-localization \cite{Pandit19,Kang20}, engineering the thermal conductance using periodically nanostructured phononic crystals \cite{Zen14,Maldovan15,Maire17}, and by compensation of the thermal drift by electrostatically induced pre-stresses on geometric anti-spring structures \cite{Zhang21}. In addition to temperature drifts, there is also a pressure dependent linear drift that is is less problematic since it can be corrected via software \cite{Middlemiss16,Middlemiss17}.

In this work, we present three strategies for a MEMS gravimeter with reduced sensitivity to temperature variations $\Delta T$. The first one is to use materials where the thermal dependence of the expansion and that of the Young's modulus cancel each other (section \ref{secCompensationTemp}). The second uses the expansion to exert a horizontal force along the beams, compensating the spring force variation (section \ref{secpressure}). The third varies the mounting point for the sensor so that the vertical expansion moves the mass back to the original position, keeping the measurement unchanged (section \ref{secexpansion}). All the strategies use a DC configuration. In appendix A we analyze the fundamental sensitivity limits due to the temperature of the resonator.

\section{\label{Theory} Mechanical gravimeter operation}
\begin{figure}
    \centering
    \includegraphics[width=\linewidth] {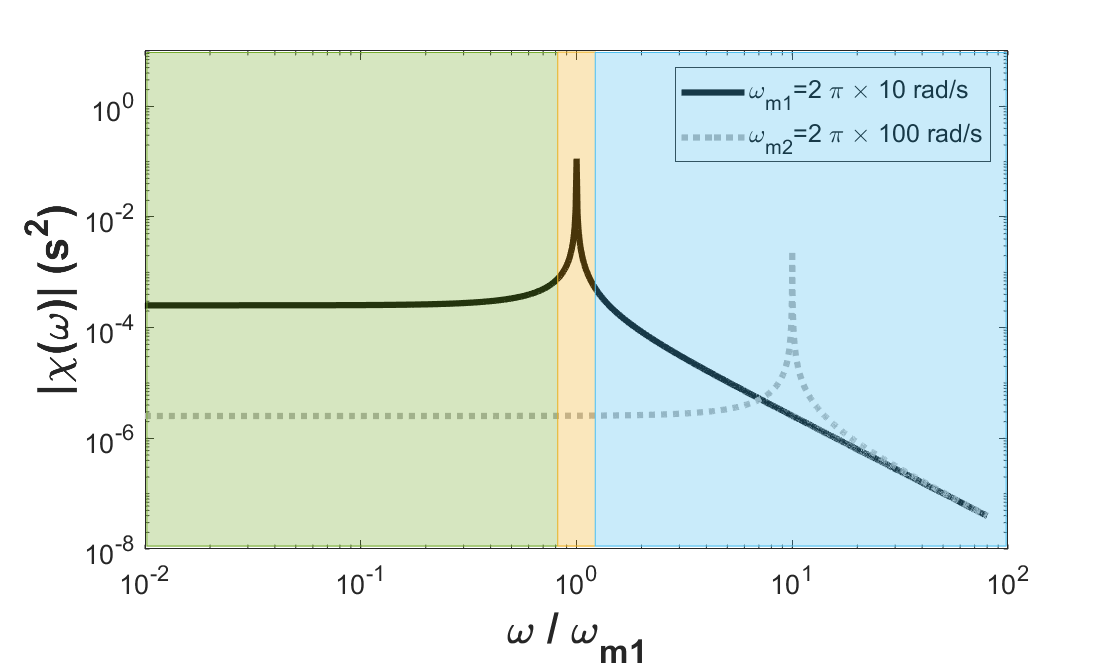}
    \caption{The magnitude of the susceptibility $\left| \chi(\omega) \right|$ as a function of angular frequency $\omega$. The solid black curve represent a gravimeter with $\omega_{m1}= 2 \pi \times 10$ rad/s and dashed gray curve with $\omega_{m2}=2 \pi \times 100$ rad/s, both with $Q_m=1000$.}
    \label{fig1}
\end{figure}

Consider a mass $m$ attached to two springs giving an effective constant $k=k_1+k_2$ in the vertical $z$ direction. The mass position is described by
\begin{equation}
\frac{d^{2}z(t)}{dt^{2}}+\gamma \frac{dz(t)}{dt}+\omega _{m}^{2}z(t)= \frac{F_{ext}(t)}{m},
\label{eq:diffEquation}
\end{equation}
with a harmonic external force $F_{ext} = \widehat{F}e^{i\omega t}=F_0e^{i(\omega t-\delta_f)}$ and $z=\widehat{z}e^{i\omega t}=z_0e^{i(\omega t-\delta_z)}$, where $F_0$ and $z_0$ are real numbers, and $\delta_F$ and $\delta_z$ the phases. The solution is
\begin{equation}
\hat{z} = \frac{\hat{F}}{m (\omega^{2}-\omega_{m} ^{2}-i \frac{\omega _{m}\omega }{Q_m})}= \chi (\omega) \hat{a},
\label{eq:susceptibility}
\end{equation}
where $\hat{a}=\hat{F}/m$ is the acceleration, and $\chi (\omega)= (\omega^{2}-\omega_{m} ^{2}-i \frac{\omega _{m}\omega }{Q_m})^{-1}$ is the transfer function or mechanical susceptibility, that gives the sensitivity of a DC gravimeter at $\omega=0$, taking $\hat{a}=g$ the local gravitational acceleration, Eq. (\ref{eq:g_relative}). There are three regimes determined by $\chi(\omega)$. For $\omega<\omega_m$ (green region in Fig. \ref{fig1}), the resonator exhibits an almost flat frequency response $\chi \simeq 1/ \omega_m^2$ and is where the DC gravimeter operates. The sensitivity is increased with a low resonant frequency $\omega_m$ (soft spring and heavy mass). The region $\omega \simeq \omega_m$ (narrow yellow region in Fig. \ref{fig1}), is where an AC gravimeter would look for a change in resonance frequency. At higher frequencies $\omega > \omega_m$ (blue region in Fig. \ref{fig1}), the resonator acts as a mechanical low pass filter. Here we focus the discussion on DC gravimeters.
\subsection{\label{performancemems}Sensitivity, long-term stability and range}
\begin{figure}
\centering
\includegraphics[width= \linewidth] {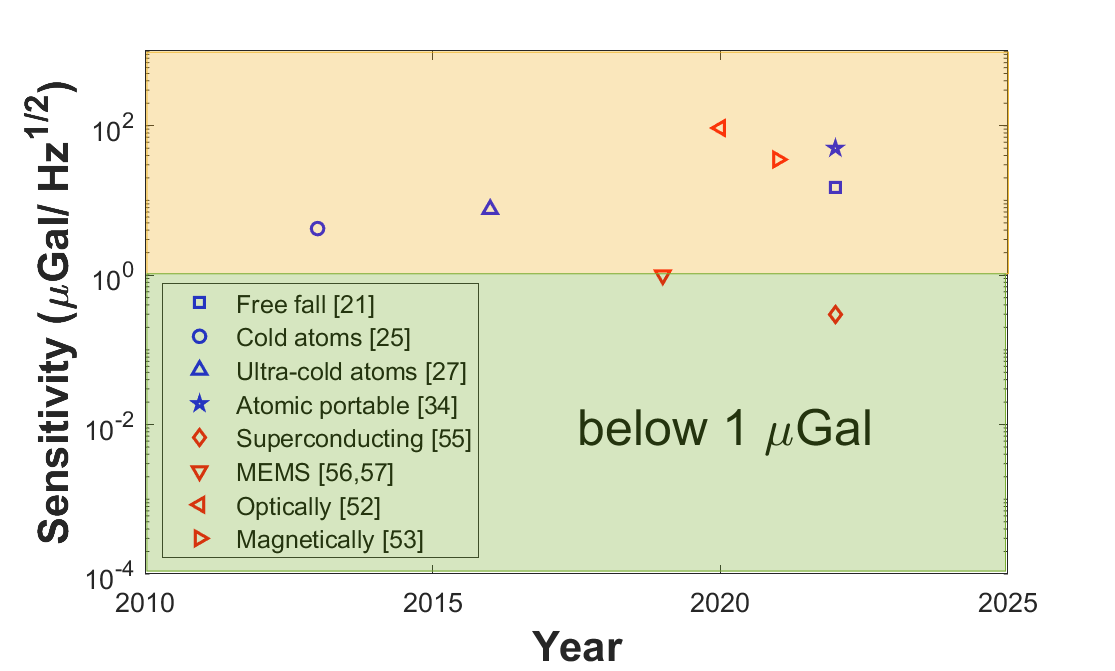}
\caption{Experimental sensitivities by both absolute (blue markers) and relative (red markers) gravimeters. The horizontal line indicates a sensitivity of $1$ $\mu$Gal/$\sqrt{Hz}$.}
\label{fig2}
\end{figure}

To detect the tiny variations of the local gravitational acceleration $g$, MEMS gravimeters must reach three goals: high sensitivity, long-term stability and wide range. Figure \ref{fig2} shows the sensitivity achieved by both absolute and relative gravimeters. It includes many of the technologies available and the sensitivity that they have reached.
The main limitation of MEMS gravimeters is the long-term stability, with an important contribution coming from temperature variations \cite{Middlemiss16,Tang19}. This includes thermal variations of the Young's modulus and the linear coefficient of thermal expansion (CTE) of silicon \cite{Bourgeois95}, which is bigger than other materials such as fused silica used in commercial gravimeters \cite{Wang20,Liu19}. In this work, we present three novel solutions to improve the long-term stability of MEMS gravimeter (section \ref{secCompensationTemp}). 

The range of MEMS gravimeters should be big enough to cover the variations of $g$ on Earth due to the location, that changes from $976$ Gal to $982$ Gal \cite{Hirt13}. It should also include the temporal variations due to Earth's tides which vary in frequency, fluctuating between diurnal ($2\times10^{-5}$ Hz) and semi-diurnal ($1\times10^{-5}$ Hz) \cite{Middlemiss16}. The mass displacement is on the scale of one millimiter, something covered with current MEMS gravimeters \cite{Middlemiss16,Tang19,Wu20,Xu21}.
\section{Thermal variations on the gravimeter}
Thermal variations affect the measurement of $g$ mainly through variations in the spring constant $k$ in Eq. (\ref{eq:g_relative}). The variation with temperature of any spatial dimension ($z$) depends on the CTE ($\alpha$) as
\begin{equation}
    \frac{\Delta z}{z} = \alpha \Delta T,
    \label{Eq:alpha0}
\end{equation}
and we write the change of the Young's modulus ($E$) with temperature as
\begin{equation}
    \frac{\Delta E}{E}=\alpha_E \Delta T,
    \label{Eq:alpha_E0}
\end{equation}
where $\alpha_E$ is the temperature coefficient of the Young's modulus.

In the section \ref{SpringTemperature} we study the thermal effects on $k$ due to $\alpha$ and $\alpha_E$. In section \ref{secCompensationTemp} we propose three ways to compensate the thermal drifts to improve the long-term stability of the MEMS gravimeter.
\subsection{\label{SpringTemperature}Temperature dependence of the spring constant}
For the simple case of a doubly-clamped beam oscillator (bridge), the resonant frequency spectrum of a rectangular beam  with uniform density $\rho$ is described by the Euler-Bernoulli Beam Theory (EBBT) for long beams ($L/t > 25$) and pure bending \cite{Yang01,Kermany17}, or by the Timoshenko Beam Theory (TBT) when the length approaches the thickness and the effect of shear and rotatory inertia is taken into account \cite{Timoshenko22,Huang61} (see Appendix \ref{Appendix:TBT}). The frequency spectrum of the flexural modes in the case of EBBT is given by \cite{Brueckner11}
\begin{equation}
f=\frac{{\kappa_n}^2 t}{4 \pi \sqrt{3}L^{2}}\sqrt{\frac{E}{\rho}}=\frac{{\kappa_n}^2}{4 \pi \sqrt{3}}\sqrt{\frac{E}{m} \eta},
\label{Eq:frequencyspectrum}
\end{equation}
where $\kappa_n = (n+\frac{1}{2}) \pi$ is the eigenvalue, $L$, $t$ and $w$ are the length, thickness and width respectively of the rectangular beam (Fig. \ref{fig3}), and $\eta =\frac{t^3}{L^3}w$ is a geometric factor. When $w \geq 25 t$, we need to replace $E$ by $E(1- \nu^2)^{-1}$ in Eq. (\ref{Eq:frequencyspectrum}), as in our case. In Appendix \ref{Appendix:TBT} we show that for the MEMS gravimeter parameters and particularly for the lower modes considered, the corrections to the spectrum due to the TBT can be neglected and it is enough to use the EBBT. The spring constant using Eq. (\ref{Eq:frequencyspectrum}) is given by 
\begin{equation}
k = m\left( 2 \pi f  \right) ^{2} =\frac{\kappa_n^4}{12} \eta E.
\label{Eq:springk}
\end{equation}

The geometric factor $\eta $ is temperature-dependent via $\alpha$ in Eq. (\ref{Eq:alpha0}). As we described in the Appendix \ref{Appendix:SiliconProperties}, silicon has an isotropic behavior of $\alpha$, therefore
\begin{equation}
    \frac{\Delta t}{t}=\frac{\Delta L}{L}=\frac{\Delta w}{w}= \frac{\Delta \eta}{\eta} = \alpha \Delta T.
    \label{Eq:alpha}
\end{equation}

Combining Eqs. (\ref{Eq:alpha_E0}), (\ref{Eq:springk}) and (\ref{Eq:alpha}), we obtain
\begin{eqnarray}
\frac{\Delta k}{k}
&=& \sqrt{\left( \frac{\Delta E}{E}\right) ^{2}+\left( \frac{\Delta \eta}{\eta}  \right)^{2}+2 \frac{\Delta E \Delta \eta}{E \eta}} \nonumber \\
&=& \left( \alpha +\alpha_E \right) \Delta T,
\label{Eq:kvsT}
\end{eqnarray}
which shows the two thermal contributions ($\alpha$ and $\alpha_E$) to the variation of $k$. Taking the force as $F=-kz$ we have
\begin{equation}
\Delta F = -k \Delta z - \Delta k z = F (\alpha_E + 2 \alpha) \Delta T.
\label{Eq:FvsT0}
\end{equation}
\subsection{\label{secCompensationTemp}Compensation of the effect of temperature variations}

Temperature fluctuations introduce a variation in the gravimeter reading, dominated in the case of silicon by the thermal change of the Young's modulus \cite{Belwanshi22n}, see Eq. (\ref{Eq:FvsT0}). Even with a temperature control of 1 mK, there is an uncertainty contribution from thermal variations of 25 $\mu$Gal \cite{Middlemiss16}. The thermal sensitivity must therefore be reduced in order to reach the 1 $\mu$Gal stability regime.

We gain some insight into the thermal sensitivity by using the analytic expression for the rectangular curved beam of Ref. \cite{Qiu04}. Considering only the lowest-order term, the force right at the center of the beam is
\begin{eqnarray}
\label{forcedisplacement}
     F & = & \left( \frac{EIh}{L^3} \right) \left[ \frac{3 \pi^4 \Theta^2}{2} \right] \Delta \left( \Delta - \frac{3}{2} + \sqrt{\frac{1}{4} - \frac{4}{3 \Theta^2}} \right) \nonumber \\
     & & \times \left( \Delta - \frac{3}{2} - \sqrt{\frac{1}{4} - \frac{4}{3 \Theta^2}} \right),
\end{eqnarray}
with $I=wt^3/12$ the moment of inertia of the beam and $h$ the initial deformation of the beam (Fig. \ref{fig3}). $\Delta=d/h$ is the displacement $d$ of the central part of the beam normalized by $h$, and $\Theta=h/t$. If $\Theta>\sqrt{16/3}$ the force changes to
\begin{equation}
    \label{f3}
    F = \left( \frac{EIh}{L^3} \right) (8 \pi^4 - 6 \pi^4 \Delta),
\end{equation}
over some range of values of $\Delta$ \cite{Qiu04}.

Temperature variations have an effect on the Young's modulus and on each linear dimension ($w$, $t$, $L$) according to Eqs. (\ref{Eq:alpha_E0}) and (\ref{Eq:alpha}) (see also Appendix \ref{Appendix:SiliconProperties} and Eqs. (\ref{Eq.TCELineal}) and (\ref{Eq:EYoungT})). If the material expansion is homogeneous, then the dimensionless quantities $\Theta$ and $\Delta$ do not contribute to the force variation, and all the thermal dependence is given by the term in the first parenthesis of Eqs. (\ref{forcedisplacement}) and (\ref{f3})
\begin{equation}
    \label{forcethermal}
    \frac{\partial F}{\partial T} = F ( \alpha_E + 2 \alpha ),
\end{equation}
which is the same as in Eq. (\ref{Eq:FvsT0}). This is a quite general result that comes from a dimensional analysis of the force, which depends linearly on E and quadratically on a spatial dimension. In the case of silicon $\alpha=2.57 \times 10^{-6}$ K$^{-1}$ \cite{Watanabe04} and $\alpha_E=-52.6 \pm 3.45 \times 10^{-6}$ K$^{-1}$ \cite{Boyd13}, so that the effect is dominated by $\alpha_E$.

Fabricating the sensor not with silicon, but with a material with $\alpha_E = -2\alpha$ should remove the dependence of the force on the temperature, that is, the effect of the change in Young's modulus would be canceled by the linear expansion. Table \ref{tab:my_label} gives a list of alloys made out of two materials that meet the above condition. Their properties change rapidly around a particular percentage of the second material. The second column gives the percentages that fulfill the above condition, where the first value we quote on each one has a less abrupt dependence on percentage, making it the more robust choice.
\begin{table}
\caption{\label{tab:my_label}Alloys that fulfil the condition $\alpha_E = -2 \alpha$ at a particular percentage of the second material. In the second column, the first percentage value would be the more robust one.}
\begin{ruledtabular}
\begin{tabular}{lcr}
Material & Percentage for $\alpha_E = -2 \alpha$ & $\alpha$ $(10^{-6})$\\
\hline
Cobalt-Palladium \cite{Masumoto70} & 90, 96  & -32.4\\
Iron-Palladium \cite{Masumoto63}  & 78, 51 & -15 \\ 
Nickel-Copper \cite{Masumoto702}  & 29, 30.5 &-29 \\
Iron-Platinum \cite{Masumoto65} & 70, 51  & -17 \\ 
Nickel-Palladium \cite{Sawaya70}& 88, 46, 59.5, 73  & 16.4 \\
Nickel-Platinum \cite{Sawaya70}& 50, 61  & 12.6\\
\end{tabular}
\end{ruledtabular}
\end{table}
Beams fabricated with the materials listed in Table 1, would give a bending behavior that remains constant with varying temperature, a quite interesting property for many applications. The problem is the availability of such alloys and the fact that they are not compatible with the same fabrication techniques as silicon, which is the material of choice for MEMS devices. In what follows we present two alternative mechanisms that can be used to enhance the effect of the linear expansion to cancel the variation due to the Young's modulus, which work with silicon and do not require any special materials. The first one takes advantage of an additional longitudinal force induced in the beam due to the expansion. The second one changes the mounting point for the frame of the sensor, to obtain a displacement of the slit with temperature, given that the gravimetry measurement relies on determining the slit position.
\subsubsection{\label{secpressure} Force from the longitudinal expansion}
\begin{figure}
\centering
\includegraphics[width=\linewidth]{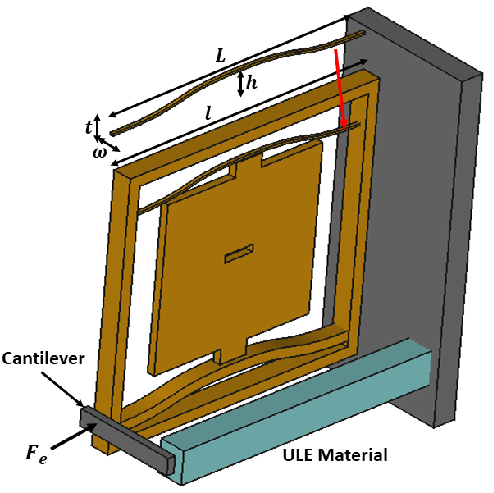}
    \caption{MEMS gravimeter (brown) with longitudinal force mechanism for temperature compensation. The blue bar is made out of a Ultra Low Expansion (ULE) material. The gravimeter and the bar are both attached to a common surface. The parameters of the beams are shown on top.}
    \label{fig3}
\end{figure}

High sensitivity on a MEMS gravimeter is achieved by having a weak spring constant $k$ around the operation point. This is achieved by combining two different springs with positive and negative stiffness respectively \cite{Tang19}. We can use Eqs. (\ref{forcedisplacement}) and (\ref{f3}) to quantify the effect of both. Consider the test mass supported by two curved beams as shown in Fig. \ref{fig3}. The upper beam is designed to have negative stiffness at the operating point, whereas the lower one has positive stiffness. We consider beams of the same width ($w$) since they would be fabricated from a single silicon wafer. By carefully adjusting the parameters of the beams, which are different for the upper and lower beam, one can obtain a total force at the  operating point (where $F=mg$) with a slope close to zero, that is, with a weak spring constant ($k=1$ N/m). We consider a test mass of $m=0.3$ g.

Applying an axial external force ($F_e$) on the lower beam that is described by Eq. (\ref{forcedisplacement}), modifies the force to
\begin{eqnarray}
\label{forcedisplacement2}
     F & = & \left( \frac{EIh}{L^3} \right) \left[ \frac{3 \pi^4 \Theta^2}{2} \Delta \left( \Delta - \frac{3}{2} + \sqrt{\frac{1}{4} - \frac{4}{3 \Theta^2}} \right) \right. \nonumber \\
     & & \times \left. \left( \Delta - \frac{3}{2} - \sqrt{\frac{1}{4} - \frac{4}{3 \Theta^2}} \right) + \frac{\pi^2 L^2}{2 E I} (1-\Delta) F_e \right].
\end{eqnarray}
We want to use this force to compensate the variation of the gravimeter reading due to temperature fluctuations. To do this, suppose we add a piece of a rigid material with no thermal expansion (for example, Ultra Low Expansion (ULE) material) parallel to the beam, as shown in Fig. \ref{fig3}. On one side both pieces are attached to a fixed common surface, while on the other there is a cantilever fixed to the ULE material that is barely touching the gravimeter. At the operating temperature, the cantilever exerts no force, but as the temperature increases, the beam expands by $\Delta l = l \alpha \Delta T$ while the ULE material maintains a constant length, so that the cantilever exerts an axial force ($F_e$ in Eq. (\ref{forcedisplacement2})) on the beam given by
\begin{equation}
    \label{pressureT}
    F_e = \kappa \Delta l = \kappa l \alpha \Delta T,
\end{equation}
with $\kappa$ the spring constant associated with the cantilever action, and $l \simeq L$ the size of the gravimeter silicon piece, assuming a thin gravimeter frame. The ULE material works as a reference length that does not expand with temperature, so that the force introduced by the cantilever depends only on the expansion of the MEMS gravimeter itself.

Consider the slit at a particular position ($d_0$) and temperature ($T_0$). With a small temperature change and displacement of the slit, the force in Eq. (\ref{forcedisplacement2}) changes to
\begin{equation}
    \label{forcedT}
    F = F + \frac{\partial F}{\partial d} \Delta d + \frac{\partial F}{\partial T} \Delta T,
\end{equation}
with $\Delta d = (d-d_0)$, and equating the force to the test mass weight gives the slit displacement
\begin{equation}
    \label{slitdispl}
    \Delta d = - \frac{(\partial F / \partial T)}{(\partial F / \partial d)} \Delta T.
\end{equation}

There are now two contributions to the temperature dependence 
\begin{eqnarray}
    \label{forcethermal2}
    \frac{\partial F}{\partial T} & = & \left[ F ( \alpha_E + 2 \alpha ) + \frac{\pi^2}{2} \alpha h_2 \kappa_2 (1-\Delta_2) \right],
\end{eqnarray}
the first term corresponds to the one given by Eq. (\ref{forcethermal}), and the second one is related with the force from Eq. (\ref{pressureT}) on the lower (positive stiffness) beam. The operating point on a gravimeter for the negative stiffness beam has a displacement close to $\Delta_1=1$. Since the effect of the force scales as $(1-\Delta)$ as in Eq. (\ref{forcethermal2}), it is more effective to apply the force in the positive stiffness beam. This contribution can be made positive or negative depending on the sign of $h_2$, that is, on the direction of the prefabricated deformation. When the temperature increases, the force is reduced as in Eq. (\ref{forcethermal}) because $\alpha_E$ is negative, and the slit position gets lowered under the action of gravity. To compensate this change, the extra term must exert a positive (upwards) force, which amounts to have a positive $h_2$.

The temperature dependence is cancelled by setting the square parenthesis in Eq. (\ref{forcethermal2}) to zero, that is, by having a spring constant of the cantilever equal to
\begin{equation}
    \label{cantilevervalue}
    \kappa_2 = - \frac{2 F (\alpha_E + 2\alpha)}{\pi^2 \alpha h_2 (1-\Delta_2)}. 
\end{equation}
A typical (silicon) beam would end up requiring a $\kappa_2$ around 100 N/m, which is about the same as that of an aluminum cantilever of length of 1 cm, width of 1 mm and thickness of 0.2 mm.

The application of an axial force with a V-beam actuator can be used to tune the stiffness of the beams in the gravimeter \cite{Zhang21}. The authors in that reference proposed measuring the temperature in a feedback loop to the actuator to correct for the temperature dependence. The compensation proposed in Eq. (\ref{cantilevervalue}) has the advantage that the correction does not depend on the quality of the feedback loop, since the change in temperature of the sensor introduces the beam expansion that corrects automatically the variation in the force.
\subsubsection{\label{secexpansion}Compensation through an expansion}
\begin{figure}
\centering
\includegraphics[width=\linewidth]{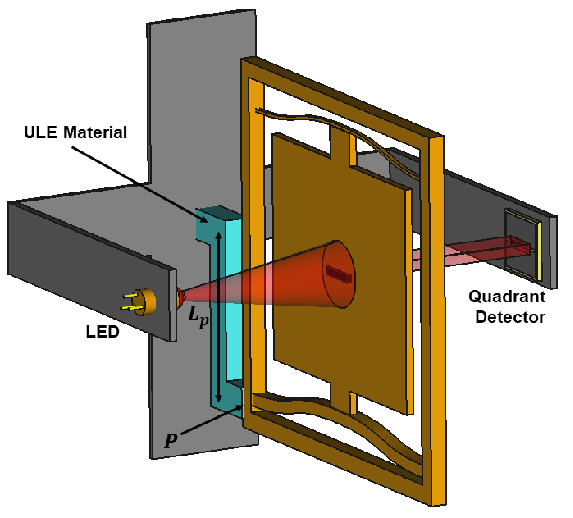}
    \caption{MEMS gravimeter (brown) with slit displacement mechanism for temperature compensation. The blue bar is made out of a Ultra Low Expansion material. The attachment point ($P$) for the sensor is at a distance $L_p$ lower compared to the height of the slit.}
    \label{fig4}
\end{figure}

Another option to correct for the temperature dependence is to apply the correction not in the force, but in the displacement of the slit in the sensor. The variation in the force introduces a slit displacement given by Eq. (\ref{slitdispl}). Taking an anchoring point for the sensor displaced vertically by a distance $L_p$ with respect to the slit, produces an slit displacement due to the thermal expansion of the sensor of $\Delta L_p = L_p \alpha \Delta T$ according to Eq. (\ref{Eq:alpha0}). The position of the slit remains unchanged with temperature variations if $\Delta d = \Delta L_p$, that is
\begin{equation}
    \label{cancelation2}
    L_p = - \frac{F}{k} \left( \frac{\alpha_E +2 \alpha}{\alpha} \right), 
\end{equation}
with $k$ the spring constant of the sensor and $F=mg$. For our parameters we obtain $L_p=5.4$ cm. Figure \ref{fig4} shows a more detailed drawing of the proposed configuration. The slit, the LED and quadrant detector they all lie at the same height. A bar of a non expanding material (ULE for example) is mounted at that same height, while the sensor frame is attached at the opposite end of the ULE material of length $L_p$. The height of this holding point ($P$) for the sensor would not be affected by a temperature change due to the use of an ULE material. Both slit displacements ($\Delta d$ and $\Delta L_p$) are generated by the same temperature change ($\Delta T$), which depends only on the temperature of the sensor itself. The sensor is small and has good thermal conductivity (148 WK$^{-1}$m$^{-1}$ \cite{Holland63,Glassbrenner64}), ensuring a uniform temperature throughout the sensor, and therefore a robust cancellation mechanism. Fine tuning of the correction is achieved by changing the attachment distance $L_p$.

One can reduce $L_p$ in order to keep an small detector by increasing $k$, at the expense of having less sensitivity to gravity. In other words, there would be a trade off between having more sensitivity and having temperature insensitivity. Increasing $k$ by a factor of 2, gives an $L_p$ two times smaller, at the price of having to integrate 4 times more to reach the same precision. Fortunately, as we show in Section \ref{performancemems}, these sensors have shown very good sensitivity \cite{Tang19}, making it worth while to sacrifice some of this sensitivity for stability. We emphasize that the two compensation mechanisms that we propose depend only on the sensor's temperature, made out of a single piece of silicon, rather than on the temperature of two different pieces, making them robust compensation mechanisms. Assembling the cantilever for the second strategy might be quite challenging, but implementing the third strategy seems within reach and requires carefully gluing the components at the right height. The thermal expansion of the ULE material must be small compared to that of silicon, the material of the sensor. This is the case working within a few degrees of the optimum temperature of the ULE material.
\section{Conclusions}
Temperature variations cause the main limitation for the long-term stability of MEMS gravimeters. They affect the spring constant because of two reasons: due to the thermal expansion that changes the dimensions of a particular design, and from a thermal variation in the Young's modulus. We propose three methods to reduce the temperature sensitivity. It is suppressed for materials that fullfil the condition $\alpha_E=-2\alpha$, that is, that the thermal expansion is canceled by the thermal variation of the Young’s modulus. We present several materials that meet this condition, reducing the sensitivity to temperature variations. MEMS gravimeters are usually made out of crystalline silicon, making it hard to use this previous strategy. An alternative is to use the sensor's own thermal expansion to introduce a force that cancels the variation in the spring constant due to temperature. Finally, we present a third alternative that focus on the displacement of the mass, which is what the sensor measures, rather than on the force. Adjusting the anchoring point of the sensor, it is possible to introduce a thermal expansion that compensates for the proof mass displacement due to the change in the spring constant. All these proposals are robust compensation mechanisms since they only depend on the temperature of the sensor itself.

\begin{acknowledgments}
We wish to acknowledge the financial support of COPOCYT Trust 23871, CONACYT A1-S-18696, CF G40 and PROFAPI PRO-A1-004 (UAS).
\end{acknowledgments}

\section*{Data Availability Statement}

The data that support the findings of this study are available from the corresponding author upon reasonable request.

\appendix

\section{Temperature variations from mechanical noise}
In this Appendix we analyze the sensitivity limits due to the temperature of the resonator, and the temperature variation $\Delta T$ due to different mechanical noise sources. We ignore the dissipation channels and assume that all that energy is converted into heat. This determines how big the quality factor $Q_m$ must be.

Thermomechanical noise gives a sensitivity limit for MEMS gravimeters in thermal equilibrium. Applying a stochastic force to the harmonic oscillator model described in section \ref{Theory}, results in a Brownian motion, which generates the thermomechanical displacement noise. The Nyquist theorem gives the thermal noise generated by a system in thermal equilibrium. In analogy with electricity, the thermomechanical force noise is \cite{Gabrielson93}
\begin{equation}
    \langle F^2 \rangle=4k_B T \frac{m \omega_m}{Q_m} B,
\end{equation}
where $k_B$ is the Boltzmann constant, and $B$ is the bandwidth. The thermomechanical acceleration noise is
\begin{equation}
    \langle a_{th}\rangle = \sqrt{ \frac{\langle F^2 \rangle}{m^2} } = \sqrt{4k_B T \frac{\omega_m}{mQ_m} B}.
    \label{anoisefloor}
\end{equation}
In the case of MEMS gravimeters based on a quasi-zero stiffness design, the theoretical acceleration noise floor at room temperature goes from $0.5$ $\mu$Gal/$\sqrt{Hz}$ \cite{Middlemiss16} up to $0.08$ $\mu$Gal/$\sqrt{Hz}$ \cite{Tang19}, which shows that sensitivity is not a limitation for a MEMS gravimeter. The last formula indicates that low frequency and high mechanical quality factors are always preferred. A DC gravimeter \cite{Tang19} reaches a sensitivity of about 1 $\mu$Gal with $Q_m>200$, which is not too demanding.

We now calculate the temperature variations $\Delta T$ due to mechanical noise sources. The power can be written as
\begin{eqnarray}
\label{Eq: power}
P &=& \frac{d}{dt}\left( K+U \right) +\gamma m\left( \frac{dz}{dt}\right) ^{2},
\end{eqnarray}
where $K=\frac{1}{2} m\left( dz/dt\right) ^{2}$ and $U=\frac{1}{2}m\omega _{m}^{2}z^{2}$ are the kinetic and the potential energy, respectively (the energy stored in the oscillation) and the last term gives the dissipation. Assuming a sinusoidal driven force $F_{ext}=F_0 \cos \left(\omega t \right)$, the general solution is $z\left( t\right)= z_{0} \cos
\left( \omega t+\delta_z \right)$
where $z_0=\frac{F_0/m}{\sqrt{\left(\omega_m^2-\omega^2 \right)^2}+\left( \omega_m \omega/ Q_m \right)^2}$ (\ref{eq:susceptibility}) and $\delta_z=\arctan \left( \frac{\omega_m^2-\omega^2}{\omega_m \omega /Q_m} \right)-\frac{\pi}{2}$. In steady state $\frac{d}{dt}\left\langle K+U \right\rangle=0$, and the last term in Eq. (\ref{Eq: power}) gives the energy that is dissipated into heat
\begin{eqnarray}
\left\langle P \right\rangle
= \left\langle \gamma m\left( \frac{dz}{dt}\right) ^{2} \right\rangle = \frac{1}{2} m \frac{\omega_m \omega^2}{Q_m} z_{0}^{2},
\end{eqnarray}
The temperature change $\Delta T$ per unit time $\Delta t$ is
\begin{equation}
\frac {\Delta T}{\Delta t} = \frac{\left\langle P \right\rangle}{m c(T)} = \frac{1}{2 c(T)} \frac{\omega_m \omega^2}{Q_m} z_{0}^{2},
\end{equation}
with $c(T)$ the specific heat capacity.

Under real conditions, some of the dominant mechanical noise sources may be due to acoustic, building motion or railroad noise. The acoustic noise corresponds to the dominant contribution, giving a heating rate below $5 \times 10^{-4}$ K/s taking silicon ($c(T)=710$ J/kg $\cdot$ K at a pressure of $1$ Pa \cite{Abe11}) with a conservative mechanical quality factor of $1200$ and a fundamental vibration frequency of $3$ Hz \cite{Tang19}. A temperature stabilization of about $1$ mK should be enough to overcome most of the above heating mechanisms.

\section{Thimoshenko Beam Theory \label{Appendix:TBT}}
The EBBT of flexural motion has been known to be inadequate for higher modes or when the effect of the cross-sectional dimensions cannot be neglected. For those cases, the TBT includes the effect of rotatory inertia and transverse-shear deformation. The coupled equations are \cite{Weaver91}
\begin{equation}
EI\frac{\partial ^{2}\phi }{\partial x^{2}}-KAG\left( \frac{\partial v}{\partial x}+\phi \right)-\rho I \frac{\partial ^{2}\phi }{\partial t^{2}}=0,
\label{Eq:coupled1}
\end{equation}
\begin{equation}
KAG \left( \frac{\partial ^{2}v}{\partial x^{2}}+\frac{\partial \phi }{\partial x}\right)-\rho A \frac{\partial ^{2}v}{\partial t^{2}}=0,
\label{Eq:coupled2}
\end{equation}
with $G$ and $E$ the shear and Young’s modulus, $K$, $A$ and $I$ the shear coefficient, area, and moment of inertia of the beam cross-section respectively, $v$ the transversal displacement and $\phi$ the cross-section rotation. Combining Eqs. (\ref{Eq:coupled1}) and (\ref{Eq:coupled2}) gives \cite{Franco16}
\begin{eqnarray}
EI\frac{\partial ^{2}v}{\partial x^{4}}-\rho I \left(1 +\frac{E}{ KG}\right) \frac{\partial ^{4}v}{\partial x^{2}\partial t^{2}}+\rho A \frac{ \partial ^{2}v}{\partial t^{2}} + \nonumber\\
\frac{ \rho^2 I}{KG}\frac{\partial ^{2}v}{\partial t^{4}} =0.
\label{eq:TE4}
\end{eqnarray}
Similarly, an equation for $\phi$ can be obtained.

Writing
\begin{equation}
v(x,t)=V(x) e^{i \omega t}, \hspace{2 mm} \phi (x,t)= \Phi (x) e^{i \omega t},
\end{equation}
gives
\begin{eqnarray}
\frac{\partial^4 V}{\partial x^4}+\frac{\rho \omega^2}{E} \left( 1+\frac{E}{KG} \right) \frac{\partial^2 V}{\partial x^2}+ \nonumber\\
\frac{\rho \omega^2}{E} \left( \frac{\rho \omega^2}{KG} - \frac{A}{I} \right) V=0.
\label{eq:TE4O}
\end{eqnarray}
The solutions has the form $V(x)=e^{\lambda^{\star} x}$ with
\begin{eqnarray}
\lambda_{+}^{\star 2},\lambda_{-}^{\star 2}=&&-\frac{\rho \omega^2}{2E}\left( 1+\frac{E}{KG} \right) \nonumber \\
&& \pm \sqrt{\frac{\rho^2 \omega^4}{4E^2} \left( 1-\frac{E}{KG} \right)^2+\frac{\rho \omega^2 A}{EI}},
\label{eq:roots}
\end{eqnarray}

which is related to the resonant frequencies for a given particular beam shape and boundary condition. In particular for a rectangular cross-section and a doubly-clamped beam, we have $K=\frac{2(1+\nu)}{4+3\nu}$ \cite{Cowper66}, $\frac{I}{A}=\frac{t^2}{12}$ and $G=\frac{E}{2(1+\nu)}$ \cite{Landau86}.  The frequency analysis is divided into two cases with respect to the \textit{transition frequency} $\tilde{\omega}^2=\frac{KGA}{\rho I}$, which is the value where $\lambda_+^{\star 2}$ changes from positive to negative ($\lambda_-^{\star 2}$ is always negative).

\begin{figure}
\centering
\includegraphics[width=\linewidth]{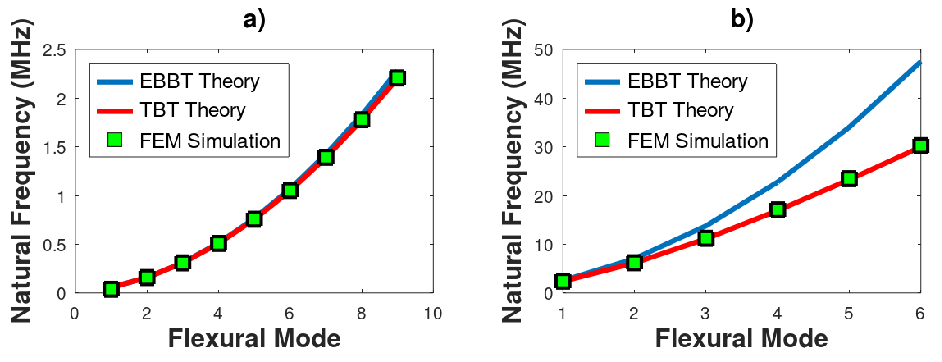}
\caption{Natural frequencies of a doubly-campled beam from the EBBT (solid blue), TBT (solid red) and finite element simulations (FEM, green square markers) for an aspect ratio of a) $L/t=10$ and b) $L/t=66$.}
\label{figA1}
\end{figure}

For the existence of non-trivial solutions, the following transcendental equation must be satisfied
\begin{eqnarray}
&&2\left( 1-\cosh \left( \lambda _{1}L\right) \cos \left( \lambda
_{2}L\right) \right) \pm  \nonumber \\
&&\frac{\lambda _{1}\lambda _{2}}{\alpha _{1}\alpha _{2}}\left( \frac{\alpha
_{2}^{2}}{\lambda _{2}^{2}}\mp \frac{\alpha _{1}^{2}}{\lambda _{1}^{2}} \right) \sinh \left( \lambda _{1}L\right) \sin \left( \lambda _{2}L\right) = 0,
\end{eqnarray}
where the upper and lower signs are for $\omega^2<\tilde{\omega}^2$ and $\omega^2>\tilde{\omega}^2$ respectively, and we have the real numbers
\begin{eqnarray}
&&\lambda_1 = \begin{cases}
+ \sqrt{+\lambda_1^{\star 2}}>0 &\text{for $\omega^2<\tilde{\omega}^2$}\\
+ \sqrt{-\lambda_1^{\star 2}}>0 &\text{for $\omega^2>\tilde{\omega}^2$}
\end{cases} \nonumber \\
&& \lambda_2=+ \sqrt{-\lambda_2^{\star ^2}}>0
\end{eqnarray}
and 
\begin{eqnarray}
&& \alpha_1 =\begin{cases}
\frac{\rho \omega^2}{KG}+\lambda_1^2&\text{for $\omega^2<\tilde{\omega}^2$},\\
\frac{\rho \omega^2}{KG}-\lambda_1^2&\text{for $\omega^2>\tilde{\omega}^2$},\\ 
\end{cases}
\nonumber \\
&& \alpha_2=\frac{\rho \omega^2}{KG}-\lambda_2^2.
\end{eqnarray}
Figure \ref{figA1} shows that for high aspect ratios, the EBBT and TBT almost gives the same result for the lower vibration modes. For small aspect ratios, EBBT and TBT start to deviate from each other, but EBBT still gives a similar result to TBT and the finite element simulations (FEM)  for the fundamental mode with deviations below $1 \%$.

\section{Temperature dependence of silicon properties \label{Appendix:SiliconProperties}}

We consider the use of single crystalline silicon (100) for the MEMS gravimeter. Table \ref{tab:siliconproperties} lists some of its properties at room temperature. The variations of the CTE and Young's modulus with temperature have been extensively studied for crystalline silicon \cite{Goeders08}. In this Appendix we model the dependence of $\alpha$ and $\alpha_E$ with temperature.
\begin{table}
\caption{\label{tab:siliconproperties}Room temperature properties of single crystalline silicon (100).}
\begin{ruledtabular}
\begin{tabular}{lr}
Property & Value\\
\hline
Density \cite{Kuramoto20} & $2330$ kg/m$^3$\\
Young´s modulus \cite{Hopcroft10,Vanhellemont14} & $130 \times 10^9$ Pa\\
Thermal conductivity \cite{Holland63,Glassbrenner64} & $148$ W/(m $\cdot$ K)\\
Poisson ratio \cite{Hopcroft10} & $0.28$\\
CTE \cite{Nye85,Lyon77} & $2.56 \times 10^{-6}$ 1/K\\
Specific heat capacity \cite{Haynes16} & $702$ J/(kg $\cdot$ K)\\
\end{tabular}
\end{ruledtabular}
\end{table}
\subsection{Linear coefficient of thermal expansion, $\alpha$ \label{TCESection}}
\begin{figure}
    \centering
    \includegraphics[width=\linewidth]{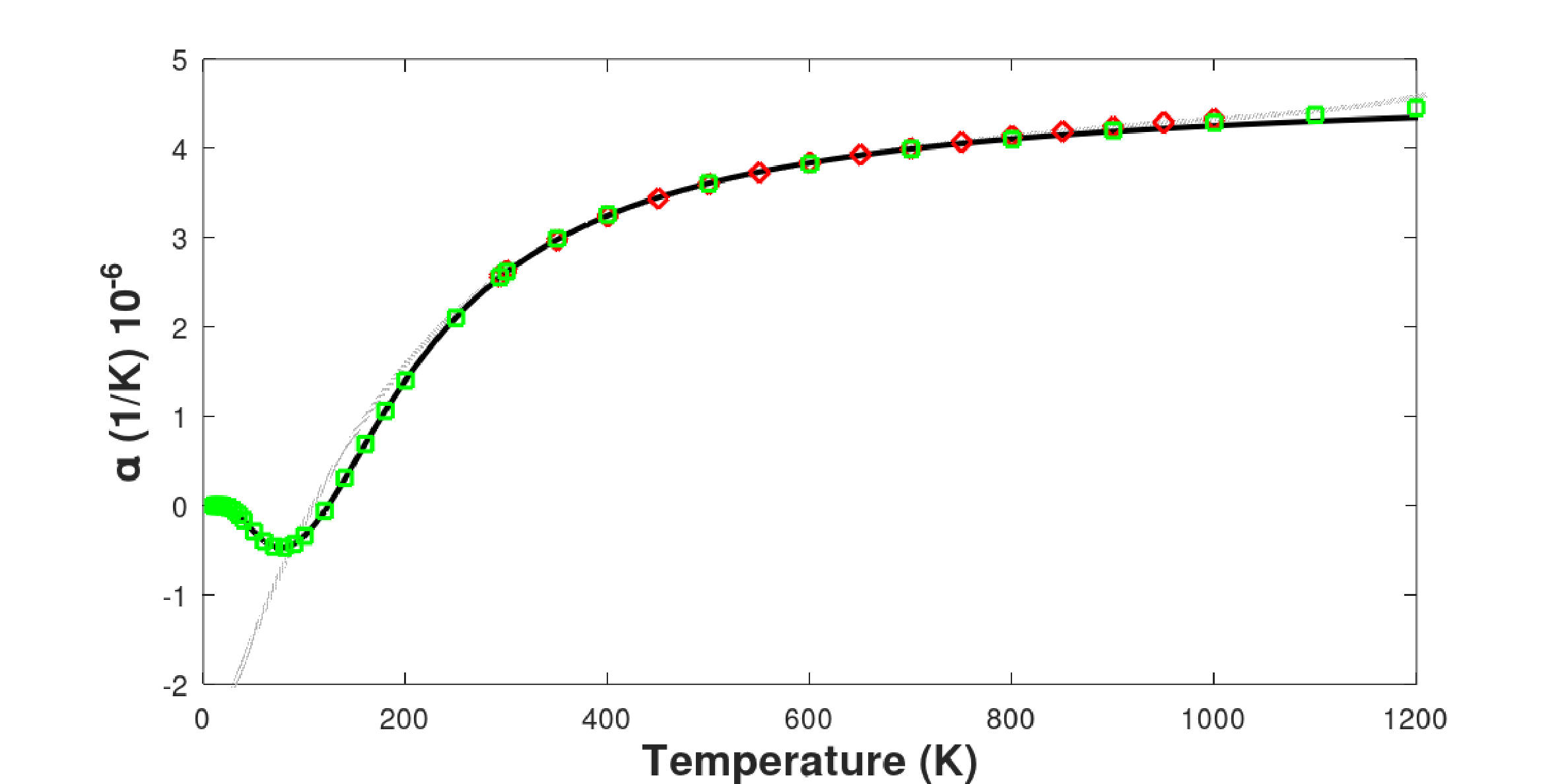}\caption{Compilation of experimental data taken by other groups for $\alpha(T)$ for $T>T_0$ (hollow red diamonds) fitted by Eq. (\ref{Eq:alfaPoly}) (dashed gray line) \cite{Watanabe04}, and for the full temperature range (hollow green squares) \cite{White97} fitted by Eq. (\ref{fullfit}) (black solid line) \cite{Swenson83CODATA,NISTsilicon}.}
    \label{figB1}
\end{figure}

The linear coefficient of thermal expansion $\alpha$, at some temperature $T$, is defined as the change in length $L(T)$ with respect to the length $L_0$ at some fixed temperature $T_0$, usually taken as the room temperature $T_0=293.15$ K \cite{Slack75}
\begin{equation}
    \alpha=\frac{d(\ln L)}{dT} \approx \frac{1}{L_0} \frac{dL}{dT}.
\end{equation}
The approximate Eq. (\ref{Eq:alpha0}) works very well because usually $L/L_0-1$ is very small for all temperatures. In all cubic crystals $\alpha$ is a scalar, independent of the direction \cite{Watanabe04,Slack75}. In single-crystal silicon a $5^{th}$ order polynomial (dashed gray line in Fig. \ref{figB1}) gives a good fit to the experimental data (hollow red diamonds) in the range from $293$ K to $1000$ K \cite{Watanabe04,Roberts81}, that is 
\begin{eqnarray}
\frac{\alpha \left( T\right) }{10^{-6} \rm{K} ^{-1}}
=&&a_0+a_1T+a_2T^{2} +a_3T^{3}+a_4T^{4}  \nonumber \\
&&+a_5T^{5},
\label{Eq:alfaPoly}
\end{eqnarray}
where $a_0=-3.0451$, $a_1=0.035705$, $a_2=-7.981\times 10^{-5}$, $a_3=9.5783\times 10^{-8}$, $a_4=-5.8919\times 10^{-11}$ and $a_5=1.4614\times 10^{-14}$. This fit does not work well for $T \leq T_0$. A good fit to the data in range from $0$ K to $600$ K (hollow green squares in Fig. \ref{figB1}) is given by a more complicated expression for  (solid black line) \cite{Swenson83CODATA,White97,Middelmann15,NISTsilicon}:
\begin{widetext}
\begin{eqnarray}
\frac{\alpha \left( T\right) }{10^{-6} \rm{K}^{-1}}  = && \left(b_0T^{3}+\left(b_1T^{5}+b_2T^{5.5}+b_3T^{6}+b_4T^{6.5}+b_5T^{7} \right) \left( \frac{1+ \rm erf \left(T_1\right) }{2}\right) \right) \left( \frac{1- \rm erf \left( 0.2 T_2 \right) }{2}\right) \nonumber \\
&&+\left( \left( -b_6+b_7 T_3^{2}+b_8 T_3^{3}+b_9 T_3 ^{9}\right) \left( \frac{1+ \rm erf \left(0.2 T_2\right) }{2}\right) \right)
 \left( \frac{1- \rm erf \left( 0.1 T_4 \right) }{2}\right) \nonumber \\
&& +\left( b_{10}+\frac{b_{11}}{T}+\frac{b_{12}}{T^{2}}+\frac{b_{13}}{T^{3}}\right)
 \left( \frac{1+ \rm erf \left( 0.1 T_4 \right) }{2}\right),
\label{fullfit}
\end{eqnarray}
\end{widetext}
where $b_0=4.8\times10^{-5}$, $b_1=1.00500\times10^{-5}$, $b_1=1.00500\times10^{-5}$, $b_2=-5.99688\times10^{-6}$, $b_3=1.25574\times10^{-6}$, $b_4=-1.12086\times10^{-7}$, $b_5=3.63225\times10^{-9}$, $b_6=-47.6$, $b_7=2.67708\times10^{-2}$, $b_8=-1.22829\times10^{-4}$, $b_9=1.62544\times10^{-18}$, $b_{10}=4.72374\times10^{2}$, $b_{11}=-3.58796\times10^{4}$, $b_{12}=-1.24191\times10^{7}$, $b_{13}=1.25972\times10^{9}$, $T_1=T-15$, $T_2=T-52$, $T_3=T-76$ and $T_4=T-200$.

\begin{figure}
    \centering\includegraphics[width=\linewidth]{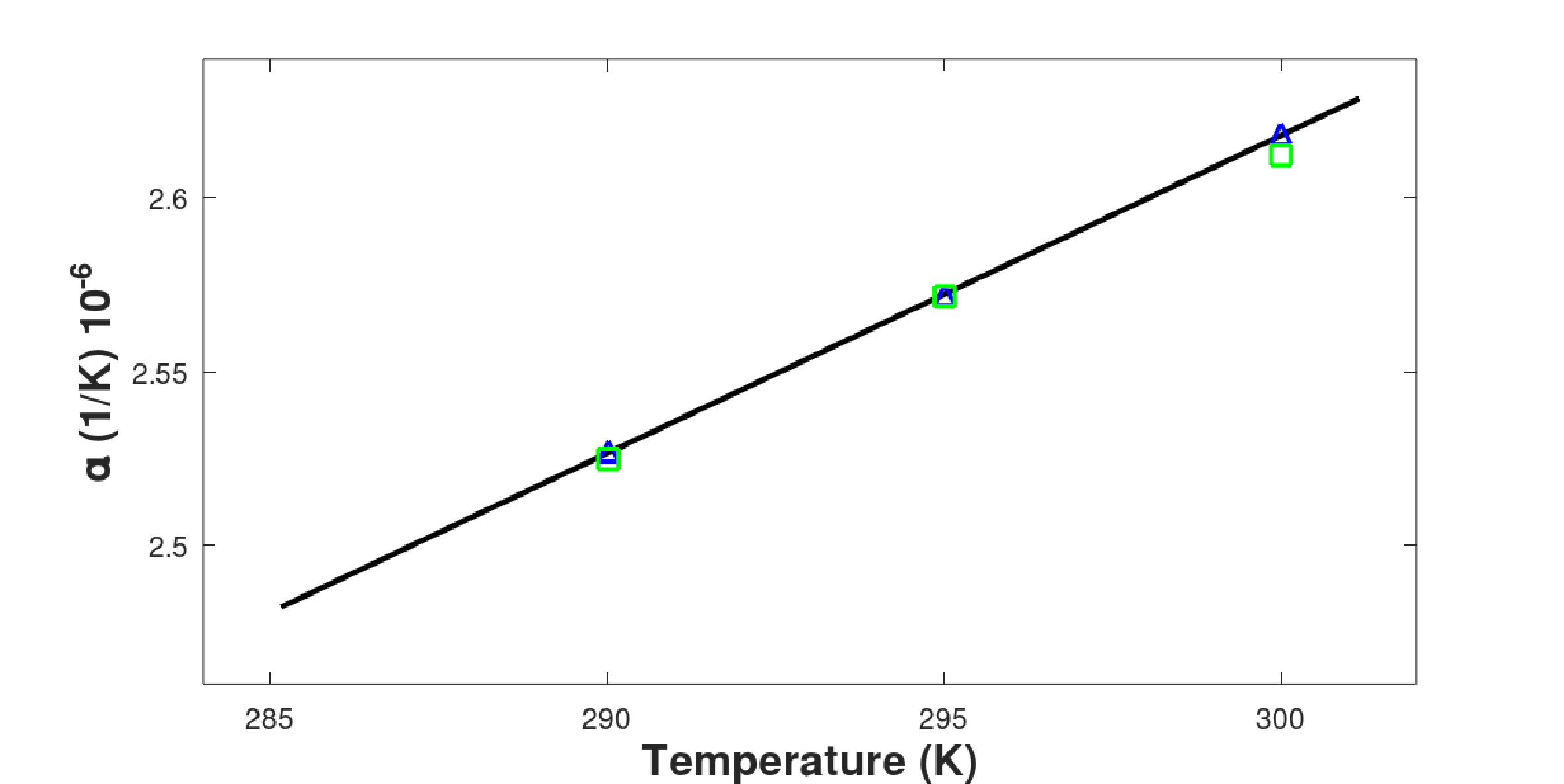}
    \caption{Compilation of experimental data taken by other groups for $\alpha(T)$ around $T_0$ \cite{Schoedel01,Bartl20} and linear fit by Eq. (\ref{Eq.TCELineal}).}
    \label{figB2}
\end{figure}

For small variations around room temperature $T_0$ a linear approximation is good enough 
\begin{equation}
    \alpha (T)=\alpha_0+2\alpha_1(T-T_0),
    \label{Eq.TCELineal}
\end{equation}
with $\alpha_0=2.5554\times10^{-6}$ K$^{-1}$ and $\alpha_1=4.58\times10^{-9}$ K$^{-2}$ (Fig. \ref{figB2}). This approximation has been verified experimentally from $285.15$ K to $301.15$ K \cite{Lyon77,Schoedel01,Karlmann06,Bartl20} (hollow blue triangles and hollow green squares in Fig. \ref{figB2}). The deviation from linearity gives a relative error on $\alpha$ of around $10$ ppm \cite{Schoedel01}.
\subsection{Young's modulus temperature coefficient $\alpha_E$}
When an isotropic solid beam of length $L$ is loaded in pure tension, the tensile stress vector is given by $\sigma$ and the longitudinal strain vector is approximately $\epsilon \sim \Delta L/L$, with $\Delta L$ the length increase. Hooke's Law states that \cite{Nye85}
\begin{equation}
    \epsilon=s \sigma,
    \label{YoungDef}
\end{equation}
where $s=1/c$ is the elastic compliance constant and $c$ the elastic stiffness constant or Young's modulus $E$. 

The compliance $s_{ij}$ with $\{i,j\}=1,2,...,6$ is a matrix. For anisotropic materials possessing a cubic crystal structure such as silicon, the matrix $s_{ij}$ contains three independent elastic constants, $s_{11}$, $s_{12}$ and $s_{44}$ from which the Young’s modulus in any crystal direction $[hkl]$ can be calculated as
\begin{eqnarray}
    \frac{1}{E_{hkl}}=&&s\hspace{0.5 mm \acute{}_{11}}=s_{11}-2 \left( s_{11}-s_{12}-\frac{1}{2} s_{44} \right) \nonumber \\
    &&\times \left( l_1^2 l_2^2+l_2^2 l_3^2+l_3^2 l_1^2 \right),
    \label{Eq.SiliconE}
\end{eqnarray}
where $l_i$ are the direction cosines along the axis \cite{Wortman65,Hall67}. Usually, experimental data is given for $c_{ij}$, instead $s_{ij}$, and their relations for silicon are
\begin{eqnarray}
c_{11} = \frac{s_{11}+s_{12}}{\left(s_{11}-s_{12}\right) \left(s_{11}+2s_{12}\right)}, \nonumber \\
c_{12} = \frac{-s_{12}}{\left(s_{11}-s_{12}\right) \left( s_{11}+2s_{12}\right)}, \qquad c_{44} &=& \frac{1}{\left( s_{44}\right)}.
\end{eqnarray}

Using $c_{11}=165.5$ GPa, $c_{12}=63.9$ GPa and $c_{44}=79.5$ GPa for silicon at room temperature \cite{Hall67}, we get $s_{11}=7.68 \times 10^{-12}$ Pa$^{-1}$, $s_{12}=-2.14 \times 10^{-12}$ Pa$^{-1}$ and $s_{44}=12.6 \times 10^{-12}$ Pa$^{-1}$ \cite{Boyd11}.

For a particular tension direction, $E$ is defined as the ratio of the longitudinal stress to the longitudinal strain, that is, $E_{hkl}=1/s \hspace{0.5 mm}\acute{}_{11}$. For the case of tension applied in the [$100$] direction, we have $l_1$$=\cos(0)=1$, $l_2$$=\cos(\pi/2)=0$ and $l_3$$=\cos(\pi/2)=0$, then $E_{100}=1/s_{11}=130$ GPa \cite{Vanhellemont14}, the minimum value for silicon. For the [$110$] direction, i.e., the direction parallel to the major flat of a ($100$) wafer, the cosines are $l_1 = \cos(\pi/2)=0$, $l_2 = \cos(\pi/4)=1/\sqrt{2}$, $l_3 = \cos(3 \pi/4) =-1/\sqrt{2}$, then $E_{110}=169$ GPa \cite{Vanhellemont14}.
 
The temperature dependence of the Young's modulus ($\alpha_E$) should be given by the thermoelastic theory, but it is hard to obtain from a simple model over a wide temperature range. However, it possible to use a semi-empirical formula \cite{Wachtman61}
\begin{equation}
    E(T)=E_{T=0}-BT \exp{\left(-\frac{T_r}{T} \right)},
    \label{Eq:EYoungExp}
\end{equation}
with $E_{T=0}$ the value of Young's modulus at absolute zero temperature, and $T_r>0$ (related to the Debye temperature) and $B>0$ two constants. Their values for silicon are obtained experimentally to be $E_{T=0}=167.5$ GPa, $T_r=317$ K and $B=15.8$ MPa/K \cite{Boyd13}.

Using (\ref{Eq:alpha_E0}) and (\ref{Eq:EYoungExp}), we have
\begin{equation}
   \alpha_E(T)= \frac{1+\frac{T_r}{T}}{T-\frac{E_{T=0}}{B} \exp \left(\frac{T_r}{T}\right)}.
   \label{Eq:EYoungT}
\end{equation}

The variation of $E(T)$ around room temperature is very linear, giving an almost constant $\alpha_E$, with an average value for silicon in the temperature range $200-300$ K of $\alpha_E=-52.6$ ppm/K \cite{Boyd13}.

\section*{References}
\bibliography{aipsamp}

\end{document}